# Ground Truths in Suicide Research:
# The Current State of AI-Based Suicide Detection in Social Media

**Authors**

Yaakov Ophir,[1,2] Ofri Hefetz,[3] Refael Tikochinski,[1,4] Kfir Bar,[3] Shir Lissak,[5] Shulamit Grinapol,[6,7] Haya Wachtel,[6] Eyal Fruchter,[3,8] Roi Reichart[5]

[1]Ariel University, [2]University of Cambridge, [3]Reichman University, [4]University College London, [5]Technion – Israel Institute of Technology, [6]Rambam Health Care Campus, [7]Maale Hacarmel Mental Health Center, [8]ICAR: Israel's Collective Action for Resilience

**Declarations**

**Ethics, Consent to Participate, and Consent to Publish Declarations:** Not applicable. This article is based on previously published literature and does not involve human participants or identifiable personal data.

**Funding:** This work was supported by the Israel Innovation Authority through the Applied Research in Academia Program.

**Conflicts of Interest:** The authors declare no conflicts of interest.

**Availability of Data:** All materials generated and analyzed in this study are available.

## Abstract

Recent advances in artificial intelligence (AI) and social media data have led to growing optimism about the ability to detect suicide risk at scale. However, the empirical foundations of this work remain unclear. This article provides a synthesis of current research on AI-based suicide detection in social media, drawing on a recent umbrella review of 22 systematic reviews covering studies up to 2022, alongside an ongoing literature review extending the analysis to more recent work.

Across these sources, we identified 195 relevant studies, which are documented in a detailed supplementary dataset outlining their key characteristics and findings (see Supplementary Information). Analysis of these studies reveals consistent patterns, including rapid growth, concentration on a small number of platforms, reliance on textual and English-language data, and repeated use of similar datasets. Most importantly, the majority of studies rely on indirect labeling strategies that do not involve direct, individual-level validation of suicide risk. Instead, ground truth is typically inferred from observable features of online content, such as linguistic markers or community membership. As a result, the predictive task often shifts from identifying individuals at risk to classifying posts that contain suicidal or distress-related language, limiting the ability of current approaches to detect individuals who do not express such content explicitly online.

These findings suggest that current advances in model performance should be interpreted with caution. Progress in this field is likely to depend less on improving model performance and more on ensuring that model predictions meaningfully correspond to suicide risk as it is experienced in real life.

## Introduction

Suicide remains one of the leading causes of potentially preventable death worldwide, motivating sustained efforts to improve the early identification of individuals at risk. Yet for decades, progress in suicide prediction remained strikingly limited. A major meta-analysis spanning roughly 50 years of research concluded that the ability of traditional statistical approaches to predict suicidal behavior was only slightly better than chance, and that the field's established risk factors had, overall, changed little over time (Franklin et al., 2017).

More recently, however, the landscape appears to have shifted. Since the publication of that meta-analysis, a growing body of studies has reported substantially higher predictive performance, often using artificial intelligence (AI) and machine learning methods applied to data derived from social media and other digital environments (Ophir et al., 2022; Resnik et al., 2020; Schafer et al., 2021). This shift has been particularly visible within the computational mental health community, where venues such as CLPsych have played a central role in advancing the field, including through shared tasks that have provided standardized datasets and benchmark tasks for suicide risk detection (e.g., Chim et al., 2024; Macavaney et al., 2021; Zirikly et al., 2019). Taken together, these developments have created the impression that a major breakthrough in suicide risk detection may now be within reach.

Much of this optimism is linked to the convergence of two transformative developments: the rise of social media and the increasing capacity of AI systems to analyze large volumes of unstructured data (Ophir & Rosenberg, 2026). As social interactions have expanded into digital spaces, online platforms have come to function, at least in principle, as a kind of window into the psychological lives of users. Unlike traditional approaches to suicide risk assessment, which typically rely on clinical encounters or structured surveys, social media provides a large-scale, naturalistic, and often unsolicited record of individuals' thoughts, emotions, and experiences. This emerging form of "digital public expression" has been seen as offering a unique opportunity to identify signals of suicide risk that might otherwise remain hidden (Ophir et al., 2022). AI methods, in turn, make it possible to process this vast and heterogeneous stream of language in ways that traditional statistical approaches could not, for example by identifying complex patterns without requiring researchers to predefine all relevant variables in advance (Badian et al., 2023; Lissak et al., 2024).

Yet despite the rapid expansion of this literature, its empirical foundations remain difficult to evaluate. The very features that make social media data attractive, including its scale, naturalistic nature, and lack of predefined structure, also introduce substantial ambiguity regarding what is actually being measured. This ambiguity raises important questions about whether recent advances can be translated into practical tools for identifying suicide risk with sufficient accuracy: What types of data do these models rely on? How are suicidal states operationalized and labeled in practice? And what exactly do current AI systems detect when they are said to identify suicide risk?

The goal of the present article is to provide a synthesis of the current state of the literature on AI-based suicide detection in social media. This synthesis draws on two complementary sources: a large-scale umbrella review of the field (Abdelmoteleb et al., 2025) and an ongoing literature review of more recent studies. The studies included in this analysis are compiled in a detailed supplementary dataset that documents their key characteristics and findings (see Supplementary Information). Taken together, this synthesis is intended to provide a timely overview of the field at a stage where rapid methodological developments and ongoing questions call for closer examination of underlying assumptions and practices.

## Analytical Approach

Our analysis drew on two complementary sources. The first focused on a 2025 umbrella review published in the *Journal of Affective Disorders*, titled "Evaluating the ability of artificial intelligence to predict suicide: A systematic review of reviews" (Abdelmoteleb et al., 2025). We applied a multi-phase screening procedure. First, we screened the 22 reviews included in the umbrella review and identified eight that were relevant to the present analysis, namely those addressing the intersection of artificial intelligence, social media, and suicidality. We then examined the titles and abstracts of all primary studies included in these reviews to assess eligibility and remove duplicate records. Finally, we reviewed the full texts of the remaining studies to confirm their alignment with the focus of the present analysis. This rigorous screening process resulted in a final corpus of 69 primary studies examining AI-based prediction of suicide risk from social media. The key characteristics and findings of these studies were systematically extracted and organized for subsequent synthesis (see Supplementary Information).

The second source analyzed in the present overview is an ongoing systematic literature review examining empirical studies at the intersection of suicidality, artificial intelligence, and social media (protocol preregistered on OSF; Ophir, 2026). This review aims to provide a structured and critical synthesis of recent research in the field, with particular attention to how suicide risk is conceptualized, operationalized, and evaluated in practice. Studies are identified through a systematic search of major databases and screened for relevance to the three focal domains. Importantly, the review focuses on studies that explicitly describe their labeling strategies, allowing for a closer examination of the ground truth assumptions underlying model development. Data extraction emphasizes key dimensions such as data sources, annotation procedures, outcome definitions, and model evaluation practices. In the present article, we report preliminary findings from studies published from 2022 onward, thereby extending the temporal scope of the synthesis beyond the period covered by the umbrella review. As of March 2026, the ongoing review has identified 126 studies meeting the inclusion criteria for the period from 2022 onwards, with only minor changes expected as the review progresses (see Supplementary Information).

## Key Findings

Across both sources, a consistent set of patterns emerges that sheds light on how current research on AI-based suicide detection in social media is conducted and what it ultimately captures.

### Field Growth

The corpus of studies identified across both sources indicates a sharp increase in research activity, with a marked acceleration in recent years. Among the 195 studies included in the supplementary dataset, fewer than half were published before 2023, while the most recent two years alone (2024 to 2026) account for approximately 43% of all studies (n = 84). This pattern reflects a broader surge of interest in computational mental health research following the widespread availability of large language models and pre-trained transformer architectures. It also highlights the importance of examining the conceptual and methodological foundations of the field at a stage of rapid expansion.

### Social Media Networks

Social media platforms are not equally represented in this literature. Reddit alone accounts for nearly half of all studies in the corpus (n = 94; 48%), making it by far the most commonly used data source. Twitter or X is a distant second (n = 46; 24%), while the Chinese microblogging platform Sina Weibo accounts for most of the remaining non-English studies (n = 19; 10%). Including studies that combine platforms, Reddit appears in 113 studies and Twitter in 64, further underscoring their central role in the field.

This concentration likely reflects a combination of methodological and structural factors. Platforms such as Reddit and Twitter provide relatively accessible data, often through public interfaces or precompiled datasets, facilitating large-scale data collection. In addition, Reddit hosts dedicated mental health communities such as r/SuicideWatch and supports longer textual posts that are well suited to linguistic analysis. At the same time, these platforms may capture only a limited segment of online expression. Other widely used platforms, including Facebook and Instagram, are more embedded in everyday social interaction and may reflect broader patterns of interpersonal and emotional life that are highly relevant in the context of suicide, yet they remain largely absent from the literature.

**Text-Based Data**

Across the full corpus, 98% of studies (n = 191) relied exclusively on textual data. Only a small number of studies incorporated visual information alongside text, and none systematically included audio, video, or behavioral metadata. This pattern stands in contrast to contemporary social media use, where image and video content increasingly constitute primary modes of self-expression, particularly among younger populations. The extent to which signals of psychological distress are expressed through non-textual modalities, and whether text-based approaches fail to capture meaningful variation in such signals, remains largely unexplored.

**Language Distribution**

Another dimension of the corpus that warrants attention is the strong concentration on English-language data. A large majority of studies (78%; n = 153) were conducted exclusively in English, with Chinese the only other language represented to a meaningful extent (n = 20; 10%). Given that suicide rates, cultural idioms of distress, and patterns of platform use vary substantially across linguistic and cultural contexts, the generalizability of models trained primarily on English-language data remains uncertain. Although it is possible that studies using other languages exist but are less visible in the predominantly English-language scientific literature, this concentration nonetheless represents an important limitation for the field.

**Dataset Reuse**

The widespread availability of Reddit-derived data has contributed to a recurring pattern across studies. A substantial proportion of the literature relies on a repeated operationalization of suicidal status based on subreddit membership, treating posts from communities such as r/SuicideWatch as indicative of suicidal ideation and posts from general communities as its absence. A particularly prominent example is a publicly available dataset of approximately 232,000 Reddit posts labeled solely on the basis of subreddit membership, which appears across multiple independent studies without modification of its labeling scheme. Beyond this case, many studies construct new datasets using the same underlying logic, effectively reproducing the same operational definition across different samples.

As a result, a growing share of the literature is based on closely related data sources and recurring definitions. Apparent improvements in model performance may therefore reflect repeated application to similar data structures rather than independent accumulation of evidence. The implications of this pattern for how suicide risk is defined and measured are considered in the final finding.

**Ground Truth**

Perhaps the most consequential finding to emerge from this synthesis concerns how suicidal states are operationalized and labeled across studies. A systematic examination of labeling practices indicates that most studies rely on indirect proxies rather than externally validated assessments of individuals. Expert annotation of posts, in which clinicians or trained raters evaluate textual content, was employed in a substantial proportion of studies. While more rigorous than simple proxy labeling, this approach remains a judgment about language rather than about the psychological state of the individual who produced it.

Direct clinical validation is rare. Only 14 studies (7.2%) established ground truth through individual-level assessment. These studies include (i) those using validated self-report instruments, (ii) those relying on self-reported behavioral outcomes such as suicide attempts, (iii) a small number employing clinically rigorous tools such as the Columbia Suicide Severity Rating Scale, and (iv) a few alternative approaches using non-standard or continuous measures. Unlike the broader literature, these studies define outcomes at the level of the individual rather than the post. Model performance in this subset is consistently lower, reflecting the greater difficulty of detecting suicide risk compared to classifying patterns of online expression.

Taken together, these findings indicate that the majority of studies rely on labeling strategies that do not involve direct, individual-level validation of suicide risk. The implications of this pattern are considered in the Discussion.

## Discussion

The present synthesis provides a structured overview of the rapidly expanding literature on AI-based suicide detection in social media. Across both sources, several consistent patterns emerge, including the concentration of studies in recent years, the reliance on a limited set of social media platforms, the predominance of text-based data, and the repeated use of similar datasets and operational definitions. Taken together, these findings suggest that the apparent growth of the field has been accompanied by a degree of methodological convergence, raising important questions about the diversity and independence of the underlying evidence base.

The most consequential implication of these patterns concerns the nature of the prediction target itself. Most studies rely on labeling strategies that do not involve direct, individual-level validation of suicide risk, with ground truth inferred from observable features of online content such as self-disclosure, linguistic markers, or community membership. As a result, the predictive task often shifts from identifying individuals at risk to classifying posts that contain suicidal or distress-related language. Models may therefore learn patterns of expression associated with particular online environments rather than indicators of underlying psychological states. This concern aligns with earlier critical reviews of predictive approaches to mental health in social media, which have highlighted persistent challenges in construct validity, particularly in ensuring that computational measures correspond to clinically meaningful mental health constructs (Chancellor & De Choudhury, 2020).

This shift has direct implications for the interpretation of model performance. High accuracy in detecting labeled outcomes does not necessarily indicate successful identification of suicide risk, but may instead reflect the consistency of linguistic patterns within the data. Moreover, when ground truth is defined at the level of observable expression, models are inherently limited in their ability to detect individuals who experience suicidal ideation but do not express it explicitly online (Ophir et al., 2022). Evidence from studies that incorporate external clinical validation further suggests that suicide risk may be reflected in more subtle linguistic patterns rather than explicit expressions of suicidality, underscoring the limitations of relying solely on post-level labeling (Lissak et al., 2024; Ophir et al., 2020).

Beyond the question of ground truth, the findings also point to broader limitations in the current evidence base. The concentration of studies on a small number of platforms, the reliance on English-language data, and the near-exclusive focus on textual modalities all raise questions about the generalizability of current models. Similarly, the repeated use of shared datasets and operational definitions may contribute to an appearance of cumulative progress that does not necessarily reflect independent validation across diverse contexts. These patterns are consistent with a recent dataset-centric review in computational mental health, which highlights systematic trade-offs between scale, accessibility, and clinical validity, as well as the dominance of a limited set of data sources and labeling strategies across studies (Gong et al., 2026).

## Limitations

Several limitations of the present study should be noted. The current synthesis does not constitute a formal systematic review, and the ongoing review on which it partly relies is still in progress. As a result, the number of included studies may change, and some of the patterns described here should be considered preliminary. In addition, the analysis is limited to studies accessible through the selected sources and may underrepresent work published in other languages or venues.

## Conclusions

Future research should aim to address these gaps by expanding the range of data sources, incorporating multimodal signals, and, critically, developing approaches to ground truth that are more closely aligned with clinically validated assessments of suicide risk. Such approaches may include the use of validated instruments such as the Columbia Suicide Severity Rating Scale (C-SSRS; Posner et al., 2011) and the Beck Scale for Suicide Ideation (BSS; Beck et al., 1979), structured clinical interviews, linkage to medical records, and documentation of clinically significant outcomes such as psychiatric hospitalization or suicide attempts. Future research may also benefit from greater attention to cultural and linguistic diversity in the expression of psychological distress, given that norms surrounding emotional disclosure and self-expression differ substantially across societies and online environments. Addressing these challenges may also help clarify what current AI systems are actually capable of detecting and under what conditions they can contribute meaningfully to suicide prevention efforts. Ultimately, progress in this field will depend less on improving model performance and more on ensuring that model predictions meaningfully correspond to suicide risk as it is experienced in real life.

## Supplementary Information

The studies included in this synthesis are compiled in a detailed supplementary dataset that documents their key characteristics and findings (see: https://osf.io/4usza/overview?view_only=87e9edef5c664678b6e128f3d905c741)

## References

Abdelmoteleb, S., Ghallab, M., & IsHak, W. W. (2025). Evaluating the ability of artificial intelligence to predict suicide: A systematic review of reviews. *Journal of Affective Disorders, 382*, 525–539. https://doi.org/10.1016/j.jad.2025.04.078

Badian, Y., Ophir, Y., Tikochinski, R., Calderon, N., Klomek, A. B., Fruchter, E., & Reichart, R. (2023). Social Media Images Can Predict Suicide Risk Using Interpretable Large Language-Vision Models. *The Journal of Clinical Psychiatry, 85*(1), 50516.

Beck, A. T., Kovacs, M., & Weissman, A. (1979). Assessment of suicidal intention: the Scale for Suicide Ideation. *Journal of Consulting and Clinical Psychology, 47*(2), 343.

Chancellor, S., & De Choudhury, M. (2020). Methods in predictive techniques for mental health status on social media: a critical review. *Npj Digital Medicine, 3*(1), 43. https://doi.org/10.1038/s41746-020-0233-7

Chim, J., Tsakalidis, A., Gkoumas, D., Atzil-Slonim, D., Ophir, Y., Zirikly, A., Resnik, P., & Liakata, M. (2024). (2024). Overview of the clpsych 2024 shared task: Leveraging large language models to identify evidence of suicidality risk in online posts. Paper presented at the 177–190.

Franklin, J. C., Ribeiro, J. D., Fox, K. R., Bentley, K. H., Kleiman, E. M., Huang, X., Musacchio, K. M., Jaroszewski, A. C., Chang, B. P., & Nock, M. K. (2017). Risk factors for suicidal thoughts and behaviors: a meta-analysis of 50 years of research. *Psychological Bulletin, 143*(2), 187.

Gong, Z., Dai, C., Ma, B., Ma, M., Sharma, M., On, T. Q., Morparia, M., Enos, B., Yu, Y., & Resnik, P. (2026). *A Survey on Mental Health Datasets and Resources*

Lissak, S., Ophir, Y., Tikochinski, R., Brunstein Klomek, A., Sisso, I., Fruchter, E., & Reichart, R. (2024). Bored to death: Artificial Intelligence research reveals the role of boredom in suicide behavior. *Frontiers in Psychiatry, 15*, 1328122.

Macavaney, S., Mittu, A., Coppersmith, G., Leintz, J., & Resnik, P. (2021). (2021). Community-level research on suicidality prediction in a secure environment: Overview of the CLPsych 2021 shared task. Paper presented at the 70–80.

Ophir, Y. (2026). Systematic Review Protocol: AI-Based Suicide Detection in Social Media. *Open Science Framework,* https://doi.org/10.17605/OSF.IO/ZCW8N

Ophir, Y., & Rosenberg, H. (2026). The technological revolution in mental health: opportunities, challenges, and practical recommendations. *Humanities and Social Sciences Communications, 13*(1), 1–9. https://doi.org/10.1057/s41599-025-06441-z

Ophir, Y., Tikochinski, R., Asterhan, C. S. C., Sisso, I., & Reichart, R. (2020). Deep neural networks detect suicide risk from textual facebook posts. *Scientific Reports, 10*(1), 16685. https://doi.org/10.1038/s41598-020-73917-0

Ophir, Y., Tikochinski, R., Brunstein Klomek, A., & Reichart, R. (2022). The Hitchhiker's Guide to Computational Linguistics in Suicide Prevention. *Clinical Psychological Science, 10*(2), 212–235. https://doi.org/10.1177/21677026211022013

Posner, K., Brown, G. K., Stanley, B., Brent, D. A., Yershova, K. V., Oquendo, M. A., Currier, G. W., Melvin, G. A., Greenhill, L., & Shen, S. (2011). The Columbia–Suicide Severity Rating Scale: initial validity and internal consistency findings from three multisite studies with adolescents and adults. *American Journal of Psychiatry, 168*(12), 1266–1277.

Resnik, P., Foreman, A., Kuchuk, M., Musacchio Schafer, K., & Pinkham, B. (2020). Naturally occurring language as a source of evidence in suicide prevention. *Suicide and Life-Threatening Behavior, 51*(1), 88–96.

Schafer, K. M., Kennedy, G., Gallyer, A., & Resnik, P. (2021). A direct comparison of theory-driven and machine learning prediction of suicide: A meta-analysis. *PloS One, 16*(4), e0249833.

Zirikly, A., Resnik, P., Uzuner, O., & Hollingshead, K. (2019). CLPsych 2019 shared task: Predicting the degree of suicide risk in Reddit posts. *Proceedings of the Sixth Workshop on Computational Linguistics and Clinical Psychology,* , 24–33.